\documentclass[twocolumn,superscriptaddress, floatfix, secnumarabic,amssymb, nobibnotes, aps, prm]{revtex4-2}

\usepackage{graphicx}
\usepackage{longtable}
\usepackage{array}
\usepackage{braket}

\newcolumntype{L}{>{\centering\arraybackslash}m{2cm}}
\newcolumntype{R}{>{\centering\arraybackslash}m{1.5cm}}
\newcolumntype{K}{>{\centering\arraybackslash}m{1.3cm}}

\begin{document}


\title{Structure and lattice excitations of the copper substituted lead oxyapatite Pb$_{9.06(7)}$Cu$_{0.94(6)}$(PO$_{3.92(4)}$)$_{6}$O$_{0.96(3)}$}  \thanks{This manuscript has been authored by UT-Battelle, LLC under Contract No. DE-AC05-00OR22725 with the U.S. Department of Energy.  The United States Government retains and the publisher, by accepting the article for publication, acknowledges that the United States Government retains a non-exclusive, paid-up, irrevocable, world-wide license to publish or reproduce the published form of this manuscript, or allow others to do so, for United States Government purposes.  The Department of Energy will provide public access to these results of federally sponsored research in accordance with the DOE Public Access Plan (http://energy.gov/downloads/doe-public-access-plan).}

\author{Qiang Zhang}
\email{zhangq6@ornl.gov}
\affiliation{Neutron Scattering Division, Oak Ridge National Laboratory, Oak Ridge, Tennessee 37831, USA}

\author{Yingdong Guan}
\affiliation{Department of Physics, Pennsylvania State University, University Park, Pennsylvania 16802, USA}

\author{Yongqiang Cheng}
 \affiliation{Neutron Scattering Division, Oak Ridge National Laboratory, Oak Ridge, Tennessee 37831, USA}

\author{Lujin Min}
\affiliation{Department of Physics, Pennsylvania State University, University Park, Pennsylvania 16802, USA}

\author{Jong K. Keum}
\affiliation{Neutron Scattering Division, Oak Ridge National Laboratory, Oak Ridge, Tennessee 37831, USA}
\affiliation{Center for Nanophase Materials Sciences, Oak Ridge national Laboratory, Oak Ridge, Tennessee 37831, USA}

\author{Zhiqiang Mao}
\affiliation{Department of Physics, Pennsylvania State University, University Park, Pennsylvania 16802, USA}

\author{Matthew B. Stone}
\email{stonemb@ornl.gov}
\affiliation{Neutron Scattering Division, Oak Ridge National Laboratory, Oak Ridge, Tennessee 37831, USA}

\date{\today}

\begin{abstract}
The copper substituted lead oxyapatite, Pb$_{10-x}$Cu$_{x}$(PO$_{3.92(4)}$)$_{6}$O$_{0.96(3)}$  (x=0.94(6)) was studied using neutron and x-ray diffraction and neutron spectroscopy techniques. The crystal structure of the main phase of our sample, which has come to be colloquially known as LK-99, is verified to possess a hexagonal structure with space group $P 6_{3}/m$, alongside the presence of impurity phases Cu and Cu$_2$S. We determine the primary substitution location of the Cu as the Pb1 ($6h$) site, with a small substitution at the Pb2 ($4f$) site. Consequently, no clear Cu-doping-induced structural distortion was observed in the investigated temperature region between 10~K and 300~K. Specially, we did not observe a reduction of coordinate number at the Pb2 site or a clear tilting of PO$_4$ tetrahedron.  Magnetic characterization reveals a diamagnetic signal in the specimen, accompanied by a very weak ferromagnetic component at 2 K. No long-range magnetic order down to 10 K was detected by the neutron diffraction. Inelastic neutron scattering measurements did not show magnetic excitations for energies up to 350 meV. There is no sign of a superconducting resonance in the excitation spectrum of this material. The measured phonon density of states compares well with density functional theory calculations performed for the main LK-99 phase and its impurity phases. Our study may shed some insight into the role of the favored substitution site of copper in the absence of structural distortion and superconductivity in LK-99.
\end{abstract}




\maketitle



\section{Introduction}

Since claims of the observation of room temperature superconductivity in a copper-substituted lead apatite compound emerged in 2023 \cite{lee2023roomtemperature}, there was a rapid growth of interest in this compound.  The parent compound, lead apatite Pb$_{10}$(PO$_4$)$_6$O was proposed to become a superconductor with the substitution of copper as Cu$^{2+}$ on the Pb$^{2+}$ sites in the structure \cite{lee2023superconductor}.   Spurred by this announcement, other researchers quickly attempted to reproduce the results.  Synthesis of the compound, which has come to be colloquially known as LK-99, was found to be accompanied by an impurity phase of Cu$_2$S although no superconductivity was observed in subsequent magnetization measurements \cite{Kumar_2023}. We note that this impurity phase was observed by the original group proposing superconductivity in this compound.   Additional synthesis by other groups also observed the Cu$_2$S impurity phase along with electrical transport and magnetization measurements that are consistent with LK-99 being a semiconductor and not a superconductor \cite{LiuAdvFuncMat}. It was argued that the first-order structural phase transition in Cu$_2$S is responsible for the signal in the magnetization, which was misinterpreted as superconductivity \cite{ZHU2023}.  Growths of LK-99 with improved purity of precursor materials or single crystals were likewise unable to reproduce the proposed high-temperature superconducting phase \cite{acsomega_kumar,Puphal2023}.
Prompted by these experimental results, density functional theory (DFT) calculations of the parent compound and the copper substituted LK-99 have been performed by several groups \cite{PhysRevB.108.L121110,LAI202466,kurleto2023pbapatite,cabezasescares2023theoretical,yang2023ab,Si_2024,korotin2023electronic,griffin2023origin}.  
Si and Held predicted an insulating electronic structure in the parent compound with a reduction in unit cell volume and flat bands near the Fermi energy appearing upon substituting Cu for Pb \cite{PhysRevB.108.L121110}. These flat bands imply a large electronic density of states near the Fermi level such that correlated electron behavior may be present in the material. Nonmagnetic DFT predicts the flat bands to have both hole and electron pockets while spin-polarized DFT calculations find that Cu substitution results in a  ferromagnetic metal with 0.57 $\mu_B$ on the Cu site and a small hybridized moment of 0.06 $\mu_B$ on the nearest neighbor O site \cite{LAI202466}.  DFT calculations examining the lattice excitations of the Cu substituted compound have found that in a hexagonal structure, bands of phonons occur in the ranges  0 to 7, 11 to 13, 15 to 17 and 27 to 30 THz accompanied by an imaginary phonon mode, while in a triclinic lattice, the phonons have similar banding but without any imaginary modes \cite{cabezasescares2023theoretical}.

Although room-temperature superconductivity in LK-99 has not been reproduced and its original observation is likely a misidentification due to a non-superconducting transition in an impurity phase, theoretical calculations do suggest the potential for correlated electron behavior.  Such calculations rely heavily on details of the crystal structure and where the Cu sites are located in the substituted compound. Griffin et al. \cite{griffin2023origin} predicted that the substitution of Cu on the Pb1 ($4f$) site induces a structural distortion and plays an important role in inducing key characteristics for high-$T_{C}$ superconductivity, such as fluctuating magnetism, flat isolated d-manifold, charge and phonons.   This underscores the importance of investigating the substitution site of the copper, particularly considering the presence of two nonequivalent Pb sites in LK-99. In addition, as compared to x-ray diffraction,  neutron diffraction is imperative for obtaining more precise information about oxygen positions and occupancy, which, unfortunately, has not been pursued. Here, in addition to the x-ray diffraction experiments, we conducted for the first time a series of neutron scattering measurements designed to better understand the nature of the crystal structure, investigate the lattice excitations, and probe for any indications of magnetic long-range order, magnetic fluctuations, or a superconducting resonance of LK-99.

      
\begin{figure*}
\centering \includegraphics[width=1\linewidth]{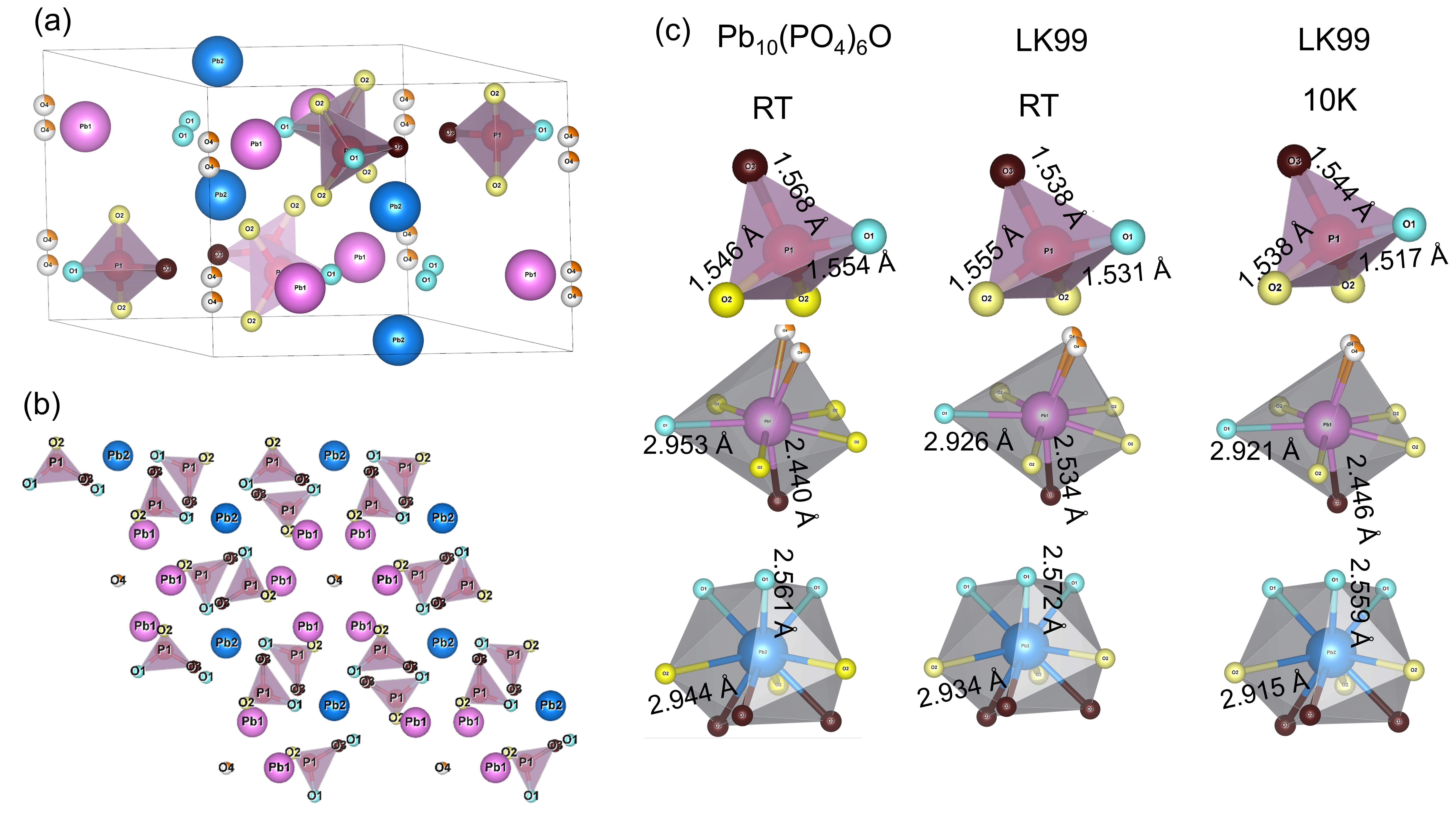}
\caption{\label{fig:crystal} 
(a) Three dimensional and (b) the projection of crystal structure in the $ab$ plane of  Pb$_{9.06(7)}$Cu$_{0.94(6)}$(PO$_{3.92(4)}$)$_{6}$O$_{0.96(3)}$ obtained from the Rietveld analysis on the XRD and neutron powder diffraction data. (c). Comparison of the chemical environments of PO$_{4}$, Pb1O$_{8}$, and Pb2O$_{9}$ in the Pb$_{10}$(PO$_{4}$)$_{6}$O at RT, Cu-doped LK-99 at RT and 10 K.  The bond lengths of P-O in PO$_{4}$ tetrahedron and the longest/shortest Pb-O bond lengths have been added for comparison.
}
\end{figure*}


\section{
\label{sec: Exp}
Experimental and Theoretical Details}

\subsection{Sample preparation and magnetization measurements}

We synthesized the Pb$_{10-x}$Cu$_{x}$(PO$_{4}$)$_{6}$O compound following the procedures reported by Lee et al.  \cite{lee2023superconductor} High purity PbO and PbSO$_{4}$ powder were thoroughly mixed and loaded into an alumina crucible. The crucible was then heated to 725 °C and kept at this temperature for 24 hours in air. The product of Pb$_{2}$(SO$_{4}$)O was verified by powder x-ray diffraction (XRD) analysis. Cu$_{3}$P was synthesized via heating the vacuum-sealed mixture of high-purity Cu and P powder. Our XRD measurements indicated that a tiny amount of Cu impurity remained in the Cu$_{3}$P product. Pb$_{2}$(SO$_{4}$)O and Cu$_{3}$P  were then mixed with a 1:1 molar ratio and sealed under vacuum in a quartz tube. The sealed tube was then heated to 925 °C and kept at this temperature for 15 hours. The formation of the  Pb$_{10-x}$Cu$_{x}$(PO$_{4}$)$_{6}$O phase was verified through Rietveld analysis on both XRD and neutron diffraction patterns. However, the product also contained impurity phases, including Cu and Cu$_{2}$S. Temperature and field dependence of the magnetization of the synthesized material was conducted using a commercial Superconducting Quantum Interference Device (SQUID, Quantum Design). 

\subsection{x-ray and neutron powder diffraction} 

XRD measurements were conducted on a PANalytical X’Pert Pro MPD equipped with an X’Celerator solid-state line detector at room temperature (RT). For the XRD measurements, the x-ray beam was generated at 45 kV/40 mA, and the x-ray beam wavelength was set at $\lambda$=1.5418 {\AA} (Cu K$\alpha$ radiation). The step size ($\Delta2\theta$) was 0.016$^{\rm{o}}$ and the exposure time at each step was 200 seconds. 

Neutron powder diffraction measurements were conducted at the time-of-flight neutron powder diffractometer POWGEN, located at the Spallation Neutron Source (SNS) at Oak Ridge National Laboratory  \cite{Powgen1,Powgen2}. The POWGEN automatic changer (PAC) was used to collect data at temperatures $T=10$ and $T=300$~K. Approximately 0.5~g of powder sample was sealed in a standard vanadium PAC can with helium exchange gas to ensure thermal contact with the PAC temperature stage. Two neutron detector banks with central neutron wavelengths of 0.8~\AA{} and 2.665~\AA{} were used to measure neutron powder diffraction over a wide range of $d$-spacing. To check the homogeneity of the sample and the substitution site of the Copper, the entire sample and two portions of the powder were subsequently used for XRD measurements at RT. The Rietveld analysis on XRD and neutron diffraction patterns was conducted using GSAS-II software  \cite{Toby2013}.

\subsection{Inelastic neutron scattering} 

Inelastic neutron scattering measurements were performed using the SEQUOIA instrument at the SNS at Oak Ridge National Laboratory   \cite{Granroth_2010,stonespec}.  The sample used was the same sample from the POWGEN measurements but sealed within a 6.35~mm diameter aluminum sample can in one atmosphere of helium gas to ensure good thermal contact with the sample environment cooling stage.  Measurements were performed using the three-sample changer closed cycle refrigerator (CCR) environment at the spectrometer.  An identical aluminum sample can was sealed and attached to the three-sample changer to provide a measurement of the empty-can scattering under the same instrument configurations as the sample measurements.  Unless otherwise noted, presented data have had this empty-can scattering subtracted.  Measurements were performed with the high-flux Fermi chopper set for incident energies of $E_i=2000$, 1000, 504, 249, 124, 59.4, and 29.5~meV with frequencies of 600, 600, 480, 360, 240, 180, and 120~Hz. respectively.   Measurements were also performed with the high-resolution Fermi chopper set for $E_i=11.4$~meV at 180~Hz.  All configurations were measured at $T=5$~K with sample(empty-can) measurement counting times of 5(0), 5(2.5), 5(2.5), 5(2.5), 7.5(2.5), 9.23(3.87), 7.29(2.53) and 10(2.5)  Coulombs of charge accumulated on the spallation target for the respective neutron energies listed above (from $E_i=2000$ to $11.4$~meV).   Such longer counting times were used to account for the reduction in neutron flux at higher incident energies and the relatively small sample size.  Some of these configurations were also measured at $T=300$~K.  The calculated full width at half maximum energy resolution for elastically scattered neutrons was $\delta\hbar\omega=240$, 86, 38, 18, 9.6, 4.3, 2.2, and 0.24 meV for the incident energies of $E_i=2000$, 1000, 504, 249, 124, 59.4, 29.5, and 11.4~meV respectively.  Powder inelastic neutron scattering data are binned into intensity histograms as a function of scattering angle, $2\theta$, and energy transfer, $\hbar\omega$ using the inelastic algorithms in the MANTID software package  \cite{ARNOLD2014156}.  Inelastic neutron scattering data are further binned as a function of the magnitude of wave-vector transfer, $Q$, and $\hbar\omega$ for presentation and analysis using DAVE software  \cite{dave}.

\subsection{DFT calculations} 
Density functional theory (DFT) calculations of Cu, Cu$_{2}$S, and LK-99 were performed using the Vienna Ab initio Simulation Package (VASP)  \cite{vasp}. The calculation used the Projector Augmented Wave (PAW)   \cite{paw1,paw2} method to describe the effects of core electrons, and Perdew-Burke-Ernzerhof (PBE) \cite{pbe} implementation of the Generalized Gradient Approximation (GGA) for the exchange-correlation functional. The spin-polarized calculation of LK-99 used the structure model and the Hubbard U (2.0 eV) as reported by Cabezas-Escares et al \cite{cabezasescares2023theoretical}. Phonopy \cite{phonopy} and finite displacement method was used to obtain the interatomic force constants and subsequently the phonon frequencies and modes. The OCLIMAX \cite{oclimax} software was used to convert the DFT-calculated phonon results to the simulated INS spectra. 

The structural model used for the DFT calculations is exactly the same as reported by Cabezas-Escares et al.\cite{cabezasescares2023theoretical} The unit cell contains 1 Cu, 9 Pb, 6 P, and 25 O atoms, with the single Cu atom occupying the 4f site. While this does not fully match the composition and Cu arrangement observed in our structural refinement, we show that this simpler model properly accounts for most of the phonon features observed in the inelastic neutron scattering experiment.  To account for the refined structure would require a much larger cell to describe the partial occupancies and disorder.  Barring a fundamental change in the electronic structure, the relatively minor differences between the model used for the DFT calculation and the refined structure are unlikely to cause significant changes in the phonon density of states.


\section{Results and Discussion}

\subsection{Rietveld analysis of the crystal structure}

\begin{figure}
\centering \includegraphics[width=1\linewidth]{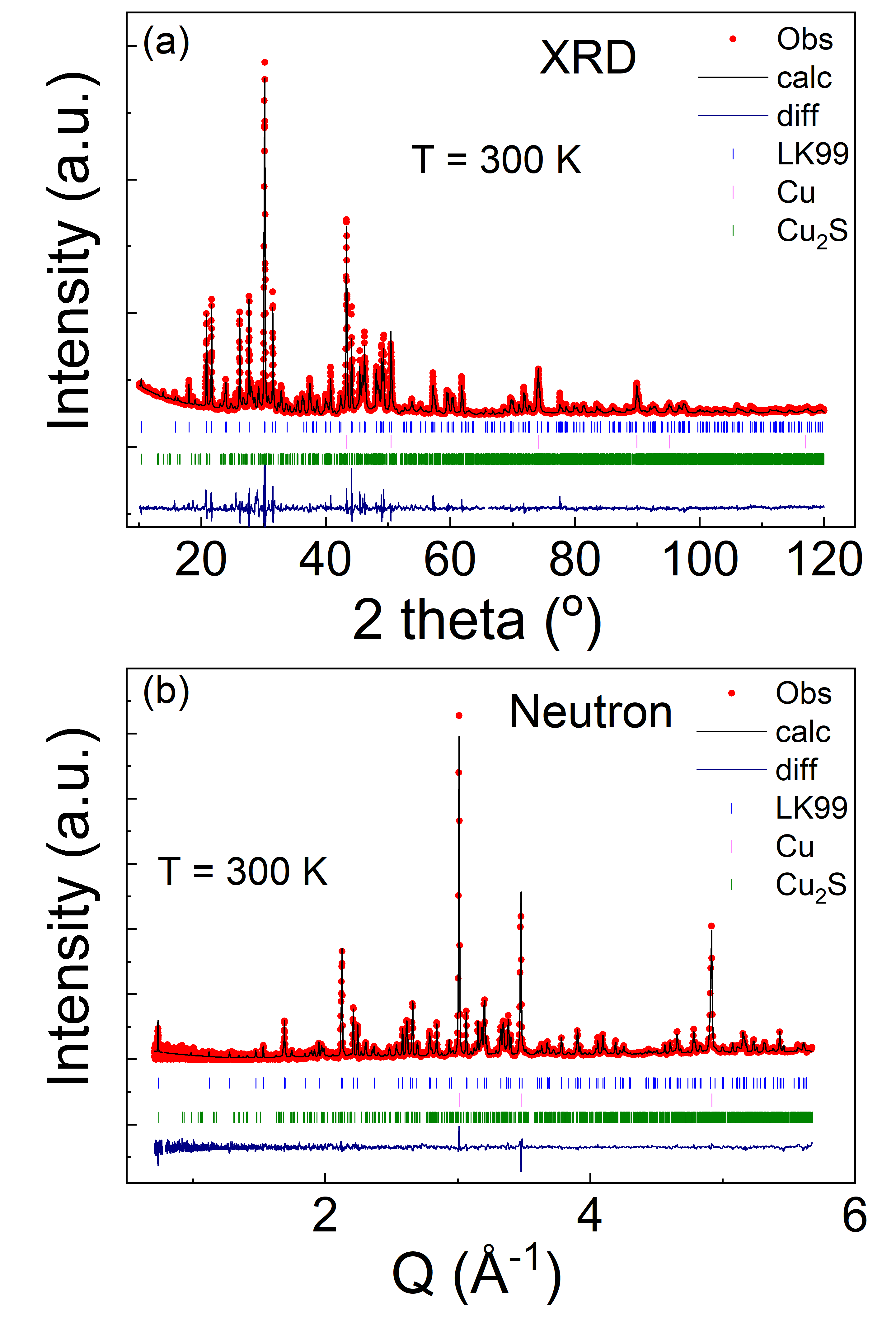}
\caption{\label{fig:DiffractionFig} 
Rietveld refinements of the (a) x-ray diffraction and (b) neutron powder diffraction patterns of  Pb$_{9.06(7)}$Cu$_{0.94(6)}$(PO$_{3.92(4)}$)$_{6}$O$_{0.96(3)}$ and its impurity phases at 300 K.
The observed data and the fit are indicated by the open circles and
solid lines, respectively. The difference curve is shown at the bottom.
The vertical bars between the data and the difference curve mark the positions of Bragg reflections for the phases
of LK-99 (upper ticks), impurity phases Cu (middle ticks) and Cu$_{2}$S (bottom ticks).
}
\end{figure}
To study the details of the crystal structure, we performed x-ray and neutron diffraction measurements on the same entire sample ($\approx$0.5 grams). The determined crystal structure is displayed in Fig. \ref{fig:crystal} (a) and (b). The XRD and neutron patterns at 300 K and the Rietveld co-refinement analysis of these patterns are shown in Fig.~\ref{fig:DiffractionFig}(a) and (b), respectively. In addition to the main LK-99 phase, there exist two impurity phases: Cu and Cu$_{2}$S. The crystal structure of Cu is cubic with space group \textit{Fm-3m} (No. 225). Cu$_{2}$S crystallizes in a monoclinic structure with a space group $P2_{1}/c$ (No. 14), consistent with previous reports  \cite{Kumar_2023,LiuAdvFuncMat,ZHU2023}.   The weight fractions of LK-99, Cu and Cu$_{2}$S are refined to be 41.2 \%, 24.1 \% and 34.8 \%, respectively, based on the co-refinement on XRD and POWGEN patterns. Rietveld analysis reveals that we have synthesized the LK-99 phase that crystallizes in the hexagonal structure with space group $P 6_{3}/m$ (No. 176). The lattice constants are refined to be $a$=9.839(7) \AA{}, $c$ = 7.428(6) \AA{} at $T=300$~K, which are consistent with the initial reports by Lee et al. on LK-99 ($a$=9.843(4) \AA{}, $c$ = 7.428(3) \AA{})  \cite{lee2023superconductor,lee2023roomtemperature}. The volume of the unit cell of our LK-99 sample is reduced by $\approx$ 0.56 \% relative to the parent compound ($a$=9.8650 \AA{}, $c$= 7.4306 \AA{})  \cite{KrivovichevBurns}, which is slightly larger than  $\approx$ 0.48\% in the previous reports by Lee et al.  \cite{lee2023superconductor,lee2023roomtemperature} due to smaller lattice constant $a$. This indicates that Pb sites may be substituted by Cu and the smaller atomic radius of Cu than Pb is responsible for the shrinking of the unit cell. Note that our Rietveld analysis on both XRD and neutron data excluded superstructures with double $c$-axis parameter as reported for the parent compound Pb$_{10}$(PO$_{4}$)$_{6}$O very recently  \cite{Sergey2023}.

Since the x-ray scattering cross-section is generally proportional to the atomic number, \textbf{Z}, the x-ray scattering factor of Pb (Z=82) is much larger than Cu (Z=29). Consequently, x-ray diffraction proves to be a powerful tool to the determination of doping concentration of Cu on the Pb1 ($6h$) and/or  Pb2 ($4f$) sites. Moreover, considering the larger neutron scattering length of Pb (9.405) in comparison to Cu (7.718), there exists an additional contrast that aids in elucidating the substitution sites of Cu by neutron diffraction. A careful Rietveld co-refinement on both XRD and neutron patterns at 300 K has discerned that the doping concentration of Cu on Pb1 and Pb2 sites is 12.9(8)\% and 3.9(6)\%, respectively.  This indicates that Cu is primarily substituted at the Pb1 ($6h$) site rather than the Pb2 ($4f$) site. Note that neutron data and XRD patterns co-refined were collected on the same 0.5 g of powder sample in its entirety. To assess  sample homogeneity and the substitution site of the Copper, two additional XRD measurements were conducted on selected partial samples. The refinement on weight fractions unveiled a non-uniform distribution of LK-99, Cu, and Cu$_{2}$S within the sample.  However, the weight fractions derived from the refinement of XRD and POWGEN patterns on the entire sample remained consistent.  Notably, despite variations in weight fractions among the three samples, Rietveld analysis on their XRD patterns yields consistent doping concentrations of Cu on the Pb1 and Pb2 sites within the uncertainties for the main LK-99 phase. This finding aligns seamlessly with the Rietveld refinement results obtained from the POWGEN data, demonstrating a robust and coherent result. Furthermore, our findings corroborate theoretical calculations  \cite{griffin2023origin}, illustrating that the replacement of Cu on the Pb1 ($6h$) site is energetically favored by 1.08 eV compared to the Pb2 ($4f$) site  (Note that the labeling of Pb1 and Pb2 in this reference [15] is reversed with our labeling). In addition, we observed negligible vacancies at P,  O1,  O2 and  O4 sites, whereas there is approximately a 7 \% vacancy at the  O3 site. The refined composition of our sample is Pb$_{10-x}$Cu$_{x}$(PO$_{3.92(4)}$)$_{6}$O$_{0.96(3)}$ (x=0.94(6)), falling into the regime of the 0.9$<$x$<$1.1 for the LK-99 in the previous reports  \cite{lee2023superconductor,lee2023roomtemperature}. 
        
 \begin{table} 
\centering
\setlength{\abovecaptionskip}{0pt}%
\setlength{\belowcaptionskip}{10pt}%
\caption{Refined atomic positions, isotropic temperature factors, occupancy, from modeling high-resolution powder neutron diffraction data of  Pb$_{9.06(7)}$Cu$_{0.94(6)}$(PO$_{3.92(4)}$)$_{6}$O$_{0.96(3)}$ at 10 K. Analysis of the nuclear Bragg reflections lead to the space group selection of $P 6_{3}/m$ (No. 176) and indexed unit cell constants of $a =b=$ 9.807(2) \AA{} and $c=$ 7.399(8) \AA{}. The refined composition is Pb$_{9.06(7)}$Cu$_{0.94(6)}$(PO$_{3.92(4)}$)$_{6}$O$_{0.96(3)}$   }
\renewcommand{\arraystretch}{1}
   \scalebox{1}{
\begin{tabular}{c|c|c|c|c|c|c}
\hline\hline
 atom & site&  x & y &  z& $U$ &  occ  \\
 \hline
Pb1  & 6h&     0.00210(5) &  0.755(3)  &   0.25000 & 0.0070(5)& 0.871(8)\\
Cu1  & 6h&     0.00210(6)  &  0.755(6)   &   0.25000 & 0.0070(5)& 0.129(8)\\
Pb2  & 4f &    0.6667  &  0.3333      &   -0.00238(4)   &  0.0070(5) &   0.961(6) \\ 
Cu2  & 4f &    0.6667 &  0.3333      &   -0.00238(4)   &  0.0070(5) &  0.039(6) \\ 
P1 & 6h &    0.624(4)  &  0.597(6)      &   0.25000   &   0.0030(7)   & 1.021(7) \\
O1 & 6h &    0.514(7) &  0.664(6)       &    0.25000   &    0.0093(8)   & 1.032(7) \\
O2 & 12i&    0.733(5)  &  0.651(2)     &  0.0842(7)   &   0.0093(8)   &  0.990(4)   \\
O3 & 6h&   0.530(6) &  0.416(8)      &  0.25000   &   0.0093(8)   &  0.933(6))   \\
O4& 4e &    0&  0    &   0.215(8)  &    0.0093(8)  & 0.241(8) \\
 \hline\hline  

\end{tabular}
 }
\label{crystal}
\end{table}
        
 It is instructive to compare the refined structural features between our LK-99 sample, the parent compound, and the prior theoretical calculations examining LK-99. In the parent compound Pb$_{10}$(PO$_{4}$)$_{6}$O  \cite{KrivovichevBurns}, three key polyhedra are noteworthy: tetrahedra PO$_{4}$, Pb1O$_{8}$ and Pb2O$_{9}$ (with the cutoff 3 \AA{} for the Pb-O bond length). Figure \ref{fig:crystal} (c) displays the comparison of these polyhedra in the parent compound at RT, our LK-99 sample at RT and 10 K based on our Rietveld refinements. The theoretical predictions by Griffin et al.  \cite{griffin2023origin} suggest that the substitution of Cu on the Pb1 ($4f$) site, corresponding to the Pb2 ($4f$) site here, induces interesting structural distortion with two primary features. The first involves a reduction in coordination around the Pb2 ($4f$) site from nine to six, presenting a distorted trigonal prism coordination. The second entails a modification and tilting of the PO$_{4}$ tetrahedra. However, in our  Pb$_{9.06(7)}$Cu$_{0.94(6)}$(PO$_{3.92(4)}$)$_{6}$O$_{0.96(3)}$  compound, the coordination of Cu/Pb1 is maintained at 9 with the same Pb-O distance cutoff of 3 \AA{}. For clarity, the longest and shortest Pb2-O bond lengths in Pb2O$_{9}$ have been included for the parent compound and our LK-99 in Fig.\ref{fig:crystal} (c). No substantial changes in the Pb2-O bond lengths or clear distorted trigonal prism were observed due to Cu doping in our LK-99 compound.  Furthermore, the coordination of Cu/Pb1 remains 8 using the same cutoff of 3 \AA{}.  The PO$_{4}$ tetrahedra in the parent compound does not form a regular tetrahedron since the O-P-O bond angle ranges from 106.4$^{\rm o}$ to 111.9$^{\rm o}$, slightly deviating from the ideal angle 109.47$^{\rm o}$. In the  Pb$_{9.06(7)}$Cu$_{0.94(6)}$(PO$_{3.92(4)}$)$_{6}$O$_{0.96(3)}$  compound, we find that the O-P-O bond angle falls in the range from 106.23$^{\rm o}$ to 111.6 $^{\rm o}$, showing comparable bond angles relative to the parent compound at RT. Thus, no clear tetrahedral tilting is observed in the LK-99 compound. The P-O bond lengths were also incorporated in Fig.\ref{fig:crystal} (c).  Both parameters suggest that the Cu doping does not induce significant tilting of the PO$_{4}$ tetrahedron. Therefore, the predicted structural distortion   \cite{griffin2023origin} was not observed in our LK-99 compound.


It comes to our attention that Griffin et al.  \cite{griffin2023origin} predicted the structural distortion under the assumption that Cu substituted to the Pb2 ($4f$) site. However, our investigation reveals a minimal substitution ratio of Cu on the Pb2 ($4f$) site, with the majority of Cu predominantly entering the Pb1 ($6h$) site in the LK-99 sample. This finding may clarify why Cu doping does not induce the predicted structural distortion in the actual LK-99 sample. Furthermore, Griffin et al.  \cite{griffin2023origin} also predicted that the substitution of Cu on the Pb2 ($4f$) site plays a pivotal role in inducing key characteristics for high-$T_{C}$ superconductivity, such as fluctuating magnetism, a flat isolated d-manifold, charge and phonons. However, the predominant incorporation of Cu into the Pb1 ($6h$) site in the LK-99 sample, as opposed to the Pb2 ($4f$) site, may offer insights into the observed absence of room-temperature superconductivity in recent experiments  \cite{cabezasescares2023theoretical}. The distinct distributions of Cu between the Pb1 and Pb2 sites necessitate a reevaluation in subsequent theoretical calculations.


\begin{figure}
\centering \includegraphics[width=1\linewidth]{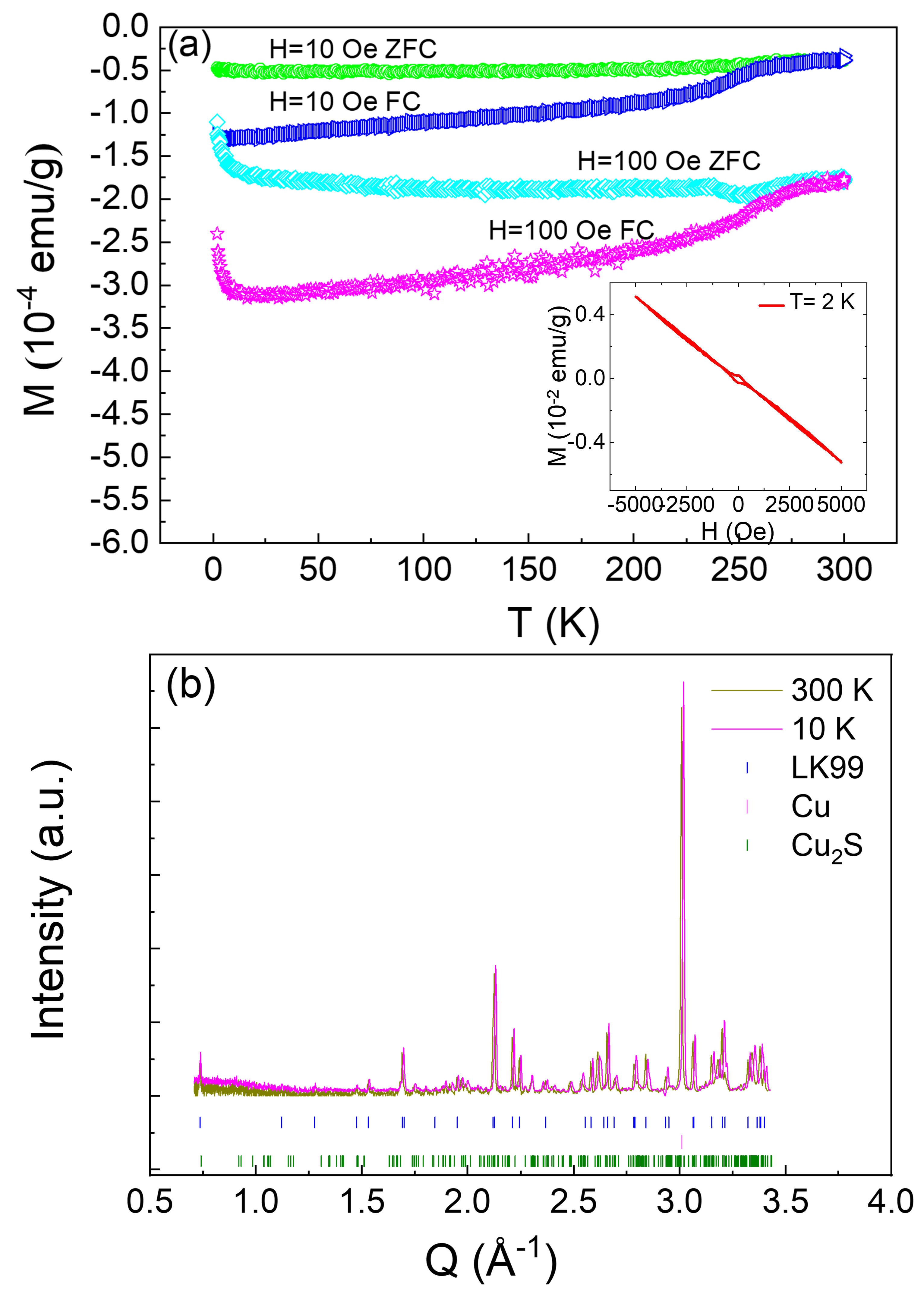}
\caption{\label{fig:mag} 
 (a) Temperature dependence of total magnetization after the ZFC and FC  processes in a low field of 10 Oe and 100 Oe of the  Pb$_{9.06(7)}$Cu$_{0.94(6)}$(PO$_{3.92(4)}$)$_{6}$O$_{0.96(3)}$ and its impurity phases. The inset shows the magnetic hysteresis loop with field sweeping from +5000 to 0 to -5000 to 0 to 5000 Oe consecutively.  (b) Comparison of low-Q POWGEN data between 300 K and 10 K showing the absence of any long-range magnetic order.  
The vertical bars between the data and the difference curve mark the positions of Bragg reflections for the phases
of LK-99 (upper ticks), impurity phases Cu (middle ticks) and Cu$_{2}$S (bottom ticks).
}
\end{figure}

\subsection{Magnetic properties} 
  The temperature dependence of the magnetization measured under zero-field-cooling (ZFC) and field-cooling (FC) histories in a field of 10 and 100 Oe is presented in Fig.~\ref{fig:mag}(a). The magnetization is negative across the investigated temperature region of 2 K to 300 K, indicative of an overall diamagnetic response from the sample. Given that the impurity phases Cu and Cu$_{2}$S are not diamagnetic in this temperature region, the diamagnetic behavior arises from the main LK-99 phase. This is consistent with observations in other polycrystalline compounds  \cite{lee2023superconductor,ZHU2023,Kumar_2023} or single crystal  \cite{Puphal2023}. The upturn observed at low temperatures in an applied field of 100 Oe implies the existence of paramagnetic impurities, a phenomenon less prominent under an applied field of 10 Oe. Note that such an upturn is much weaker than those in the previous reports  \cite{lee2023superconductor,ZHU2023,Puphal2023}, indicating a lower fraction of paramagnetic impurities in our sample.   
  The magnetic hysteresis loop,  illustrated in the inset of Fig.~\ref{fig:mag} exhibits diamagnetic behavior. A small coercive field  $\approx$ 400 Oe at 2 K indicates a very weak ferromagnetic component. This finding is consistent with observations in LK-99 single crystals  \cite{Puphal2023}, possibly arising from frustrated exchange interactions within the Cu-rich clusters.

\begin{figure}
\centering \includegraphics[width=1\linewidth]{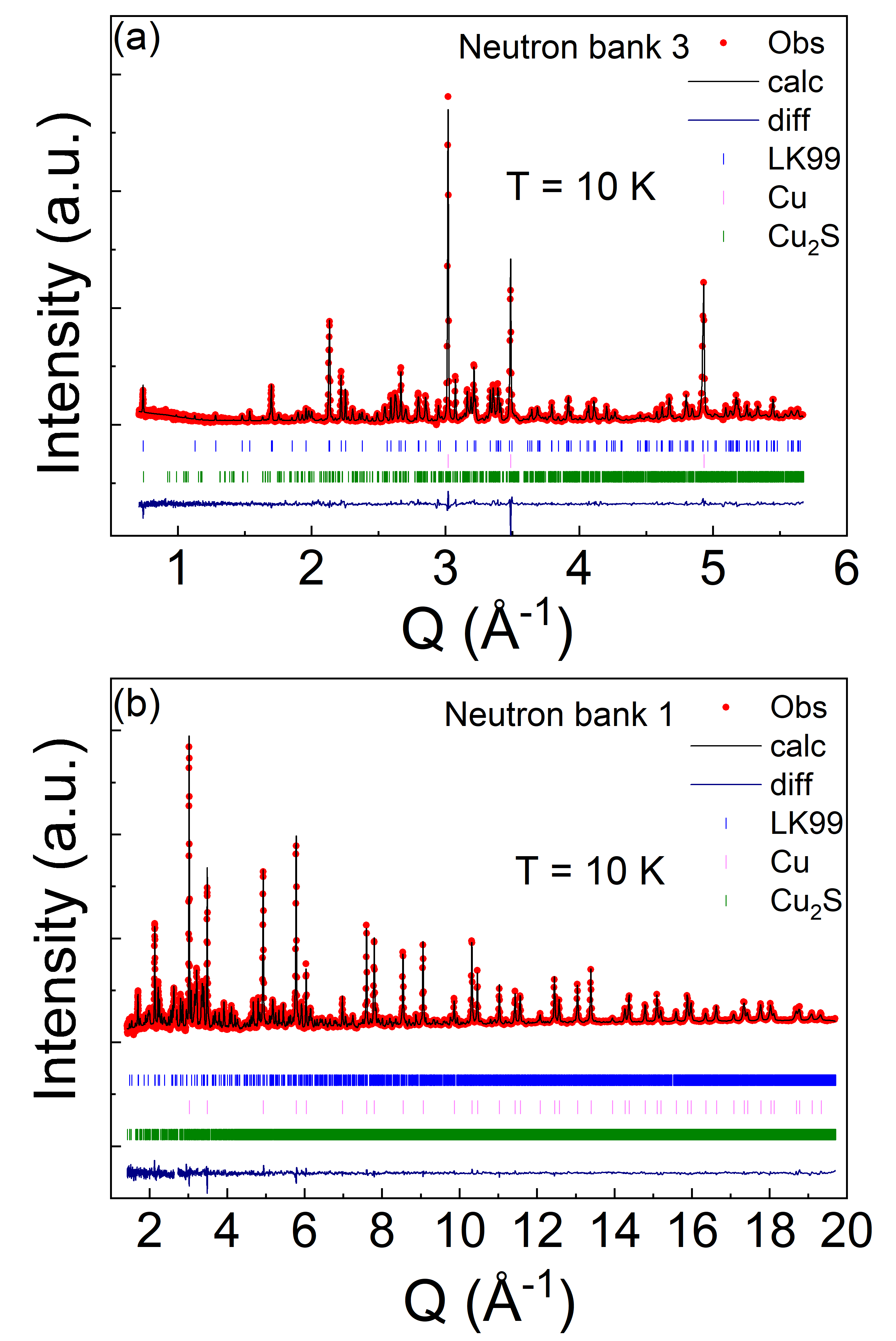}
\caption{\label{fig:10K} 
Rietveld refinement fit to neutron powder diffraction patterns collected by (a) neutron bank 3 and (b) neutron bank 1 for  Pb$_{9.06(7)}$Cu$_{0.94(6)}$(PO$_{3.92(4)}$)$_{6}$O$_{0.96(3)}$ and its impurity phases at the POWGEN instrument at 10 K.
The observed data and the fit are indicated by the open circles and
solid lines, respectively. The difference curve is shown at the bottom.
The vertical bars mark the positions of Bragg reflections for the phases
of LK-99 (top), impurity phases Cu (middle) and Cu$_{2}$S (bottom).
}
\end{figure}

  The overplot of the POWGEN data in the low-$Q$ region at 300 K and 10 K is displayed in Fig.~\ref{fig:mag}(b). No discernible pure magnetic peaks or an increase in peak intensity attributed to magnetic contributions were observed. Consequently, there is no indication of long-range magnetic order from 300 K down to 10 K in LK-99.  The POWGEN data at 10 K can be well fitted using these three phases: LK-99, Cu and Cu$_{2}$S, as shown in Fig.~\ref{fig:10K}(a-b). Furthermore, there is no structural transition or clear structural distortion (see Fig. \ref{fig:crystal}(c)) observed down to 10 K. The refined lattice constants, atomic positions, isotropic temperature factors and occupancy at 10 K are summarized in Table \ref{crystal}.

\subsection{Lattice excitations} 

\begin{figure}
\includegraphics[scale=0.75]{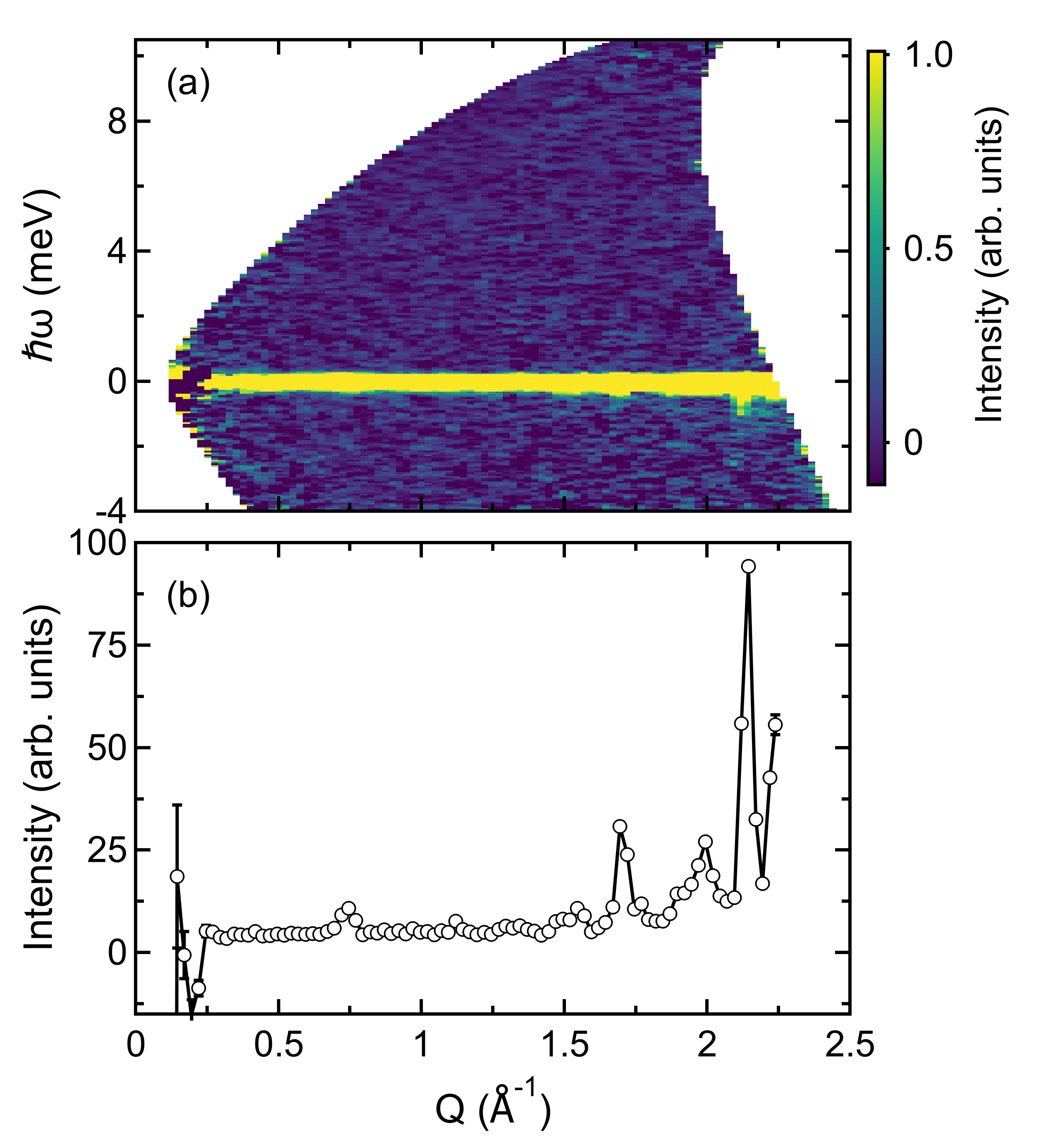} 
\caption{\label{fig:dispEi11} 
Scattering intensity of  Pb$_{9.06(7)}$Cu$_{0.94(6)}$(PO$_{3.92(4)}$)$_{6}$O$_{0.96(3)}$ and its impurity phases  at $T=5$~K measured with $E_i=11.4$~meV neutrons.  Data have been background subtracted as described in the text.  (a) Scattering intensity as a function of $Q$ and $\hbar\omega$.  (b) Scattering intensity for $-0.24<\hbar\omega<0.24$~meV as a function of $Q$.
}
\end{figure}

Figure~\ref{fig:dispEi11}(a) shows the measured scattering intensity as a function of $Q$ and $\hbar\omega$ for $E_i=11.4$~meV neutrons measured at $T=5$~K.  There are no apparent magnetic fluctuations present in the spectrum.  Figure~\ref{fig:dispEi11}(b) shows the integrated scattering intensity for elastically scattered neutrons with $-0.24<\hbar\omega<0.24$~meV.  The reflections at $Q\approx 0.74$ and $1.1$~{\AA}$^{-1}$ corresponding to d-spacings of $\approx 8.5$ and $5.7$~{\AA} respectively, are in agreement with the 010 and 011 reflections observed in the POWGEN measurements.  No additional Bragg reflections, beyond those measured at POWGEN, are discernible in the inelastic measurements.  Higher $Q$ peaks are also structural and consistent with the POWGEN measurements.

\begin{figure}[htbp]
\includegraphics[scale=0.8]{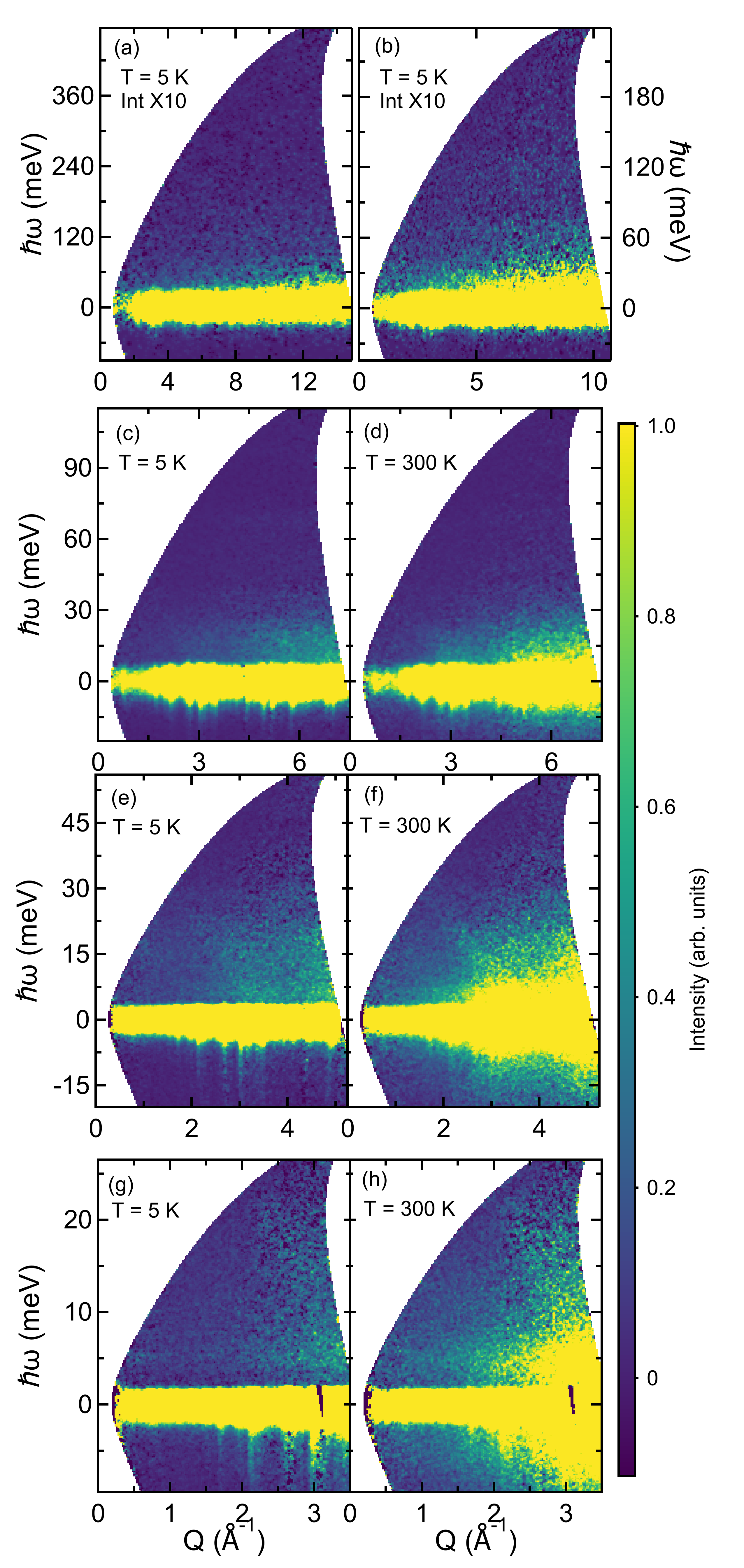} 
\caption{\label{fig:contours} 
Scattering intensity of Pb$_{9.06(7)}$Cu$_{0.94(6)}$(PO$_{3.92(4)}$)$_{6}$O$_{0.96(3)}$ and its impurity phases at $T=5$~K and $T=300$~K as a function of $Q$ and $\hbar\omega$ measured with different incident energy, $E_i$, neutrons.  Data have been background subtracted as described in the text.  (a) Scattering intensity for $Ei=504$~meV and $T=5$~K.  (b) Scattering intensity for $Ei=249$~meV and $T=5$~K.  (c)-(d) Scattering intensity for $Ei=124$~meV and $T=5$~K and $T=300$~K respectively.  (e)-(f) Scattering intensity for $Ei=59.4$~meV and $T=5$~K and $T=300$~K respectively.   (g)-(h) Scattering intensity for $Ei=29.5$~meV and $T=5$~K and $T=300$~K respectively.  The measured intensity shown in panels (a) and (b) is multiplied by a factor of 10.
}
\end{figure}

Figure~\ref{fig:contours} shows the measured scattering intensity as a function of $Q$ and $\hbar\omega$ for several incident energy measurements at both $T=5$ and 300~K. For the $Ei=504$, 249, 124, and 60~meV measurements, there is no significant scattering observed at lower values of wave-vector transfer that would be consistent with magnetic scattering.  The proposed superconducting transition temperature of Pb$_{10-x}$Cu$_x$(PO$_4$)$_6$O was $T_C\approx 400$~K  \cite{lee2023roomtemperature,lee2023superconductor}.  If LK-99 were an unconventional  superconductor, one would expect potential superconducting resonances in the excitation spectrum  \cite{superconductingresonance}.  Superconducting resonance energies, $E_r$, in inelastic neutron scattering measurements are observed at temperatures that are linearly proportional to the transition temperature.  For example in the Fe-based superconducting compounds, this relationship has been found to be $E_r=4.9k_B T_C$  \cite{PhysRevLett.125.117002}, where $K_B$ is Boltzman's constant, and $E_r=5.8k_B T_C$ for the cuprate superconductors  \cite{PhysRevLett.99.017001}.  Using the cuprate characteristic energy of the resonance for LK-99 would place the superconducting resonance at approximately 200 meV at low $Q$.  No such feature is observed in our measured spectra.  At larger values of wave-vector transfer, a powder phonon density of states (PDOS) with bands of phonons extending up to  $\hbar\omega\approx120$~meV is observed.

\begin{figure}
\includegraphics[scale=0.85]{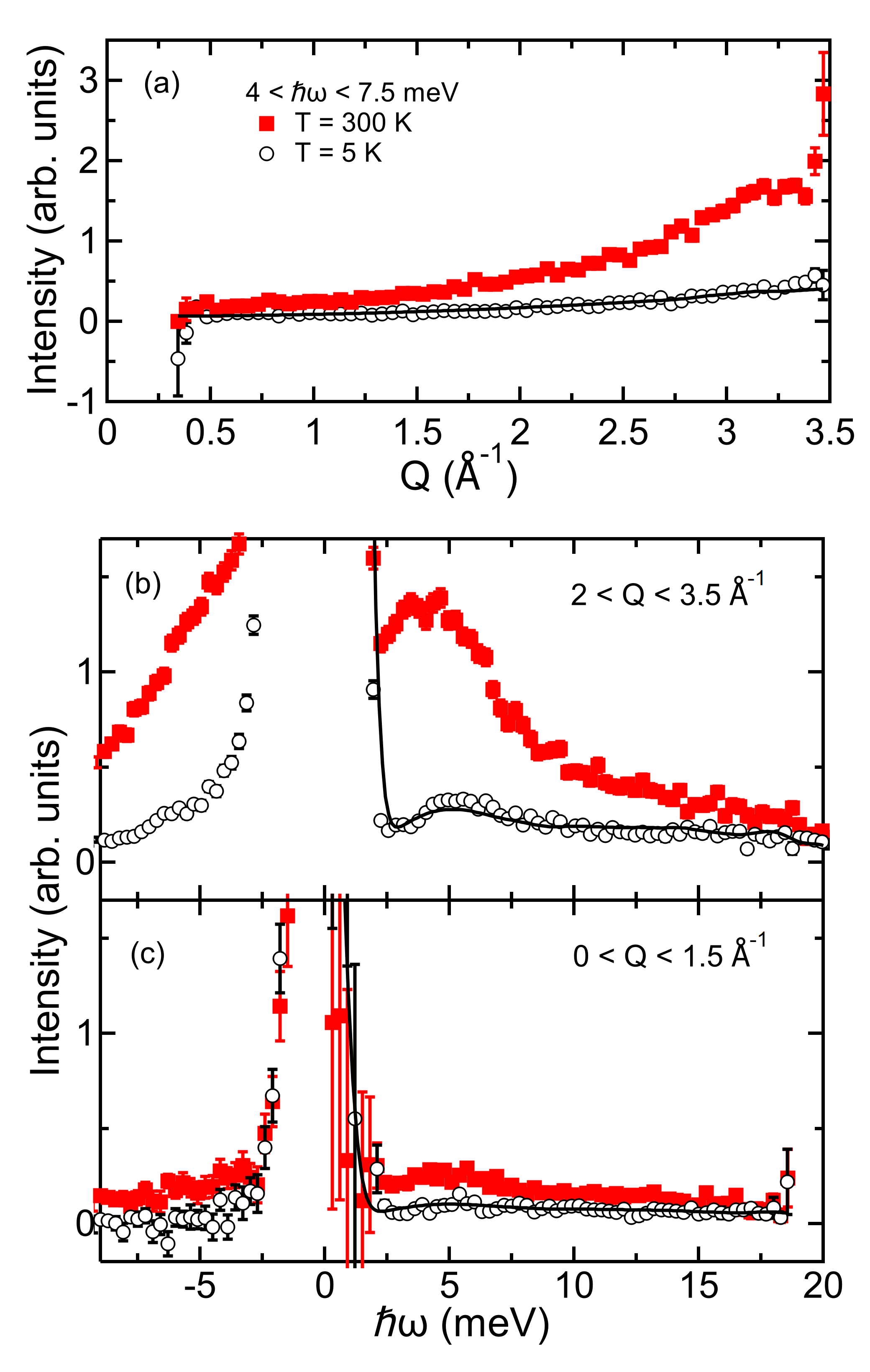} 
\caption{\label{fig:ei30cuts} 
Scattering intensity of Pb$_{9.06(7)}$Cu$_{0.94(6)}$(PO$_{3.92(4)}$)$_{6}$O$_{0.96(3)}$ and its impurity phases at $T=5$~K and $T=300$~K for the $Ei=29.5$~meV measurement.  Data are cut through the spectra shown in Figs.~\ref{fig:contours}(g)-(h) as a function of $Q$ and $\hbar\omega$.  Data have been background subtracted as described in the text.  (a) Scattering intensities as a function of $Q$ for energy transfers $4<\hbar\omega<7.5$~meV.  (b) Scattering intensities for wave-vectors $2<Q<3.5$~{\AA}$^{-1}$.  (c)  Scattering intensities for wave-vectors $0<Q<1.5$~{\AA}$^{-1}$.  Solid lines are fits to the $T=5$~K data as the sum of scattering contributions from LK-99, Cu$_2$s, and Cu as described in the text.  The range of the solid line along the independent variable indicates the range used for the fit of the data.}
\end{figure}

The measured spectra for $E_i=29.5$~meV have some scattering intensity appearing near $\hbar\omega=6$~meV. This is present in both the $T=5$~K and $T=300$~K data as shown in Figs.~\ref{fig:contours}(g) and (h), respectively.  Figure~\ref{fig:ei30cuts} are plots of scattering intensity as a function of (a) $Q$ and (b)-(c) $\hbar\omega$ for the $T=5$ and 300~K measurements.  The wave-vector dependence, Fig~\ref{fig:ei30cuts}(a) is integrated between 4 and 7.5 meV and shows quadratic behavior for both temperatures consistent with this mode being associated with a lattice excitation.  Intensity as a function of energy transfer for two different ranges of integration of wave-vector transfer, Figs.~\ref{fig:ei30cuts}(b)-(c), show that the $T=5$~K mode has a peak in the vicinity of 5.5 meV which softens slightly with increasing temperature and increases rapidly in intensity as a function of wave-vector transfer making it further consistent with it being a lattice excitation. 

\begin{figure}
\includegraphics[scale=0.85]{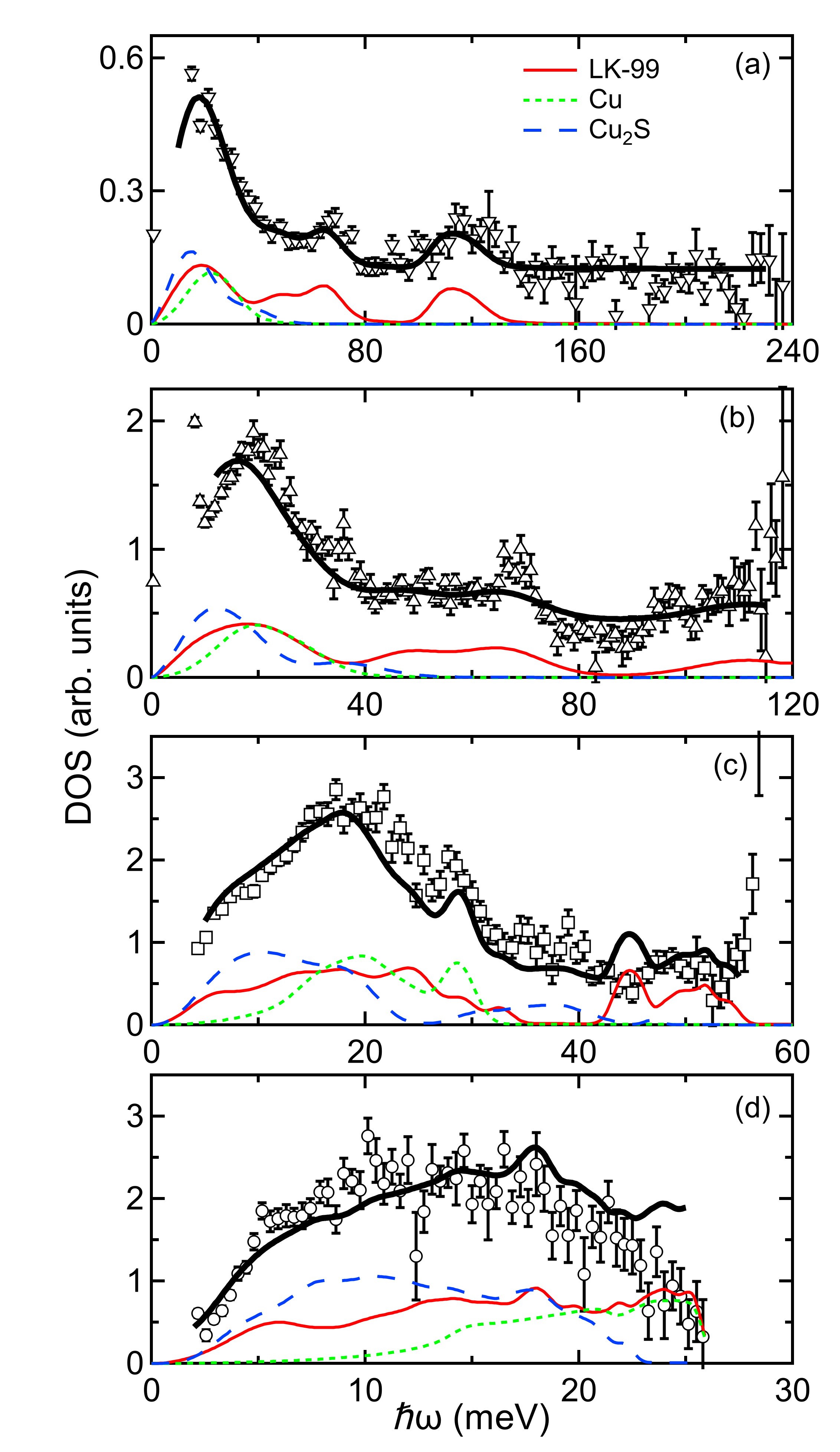} 
\caption{\label{fig:ndos} 
Generalized phonon density of states for Pb$_{9.06(7)}$Cu$_{0.94(6)}$(PO$_{3.92(4)}$)$_{6}$O$_{0.96(3)}$ and its impurity phases for $T=5$~K.  Data have been extracted into a general phonon density of states as described in the text.  Data were acquired with four different incident energy configurations as described in the text: (a) $E_i=249$~meV, (b) $E_i=124$~meV, (c) $E_i=59.4$~meV, and (d) $E_i=29.5$~meV.  The red solid lines are the calculated phonon density of states of  Pb$_{9.06(7)}$Cu$_{0.94(6)}$(PO$_{3.92(4)}$)$_{6}$O$_{0.96(3)}$  convolved with the respective energy transfer dependent resolution function for each of the measurement configurations as described in the text.  The dashed blue and green dotted lines are similar calculations for Cu$_2$S and Cu, respectively, as described in the text.   The heavy solid line is the refined combined phonon density of states that has been fit simultaneously to all of the measurements shown.  This fit is further described in the text.  The range used for the is the range of the black line.  The individual phonon density of states is shown as their respective contribution to the overall measured phonon density of states but with no additional flat background included.}
\end{figure}

We individually examine the generalized phonon density of states (PDOS) from the different incident energy neutron spectroscopy measurements.  The scattering intensity was first plotted as a function of wave-vector transfer for the $E_i=29.5$, $59.4$, and $124$~meV configurations integrated over energy transfer ranges $[4,8]$, $[10,30]$, and $[60,74]$~meV respectively.  These data were individually compared as a function of energy transfer to the function $I(Q) = A + B*Q^2 \exp(-|u|^2 Q^2)$ where $A$ is an overall constant background, $B$ is a multiplicative constant, and $u$ quantifies the Debye-Waller factor.  The mean Debye-Waller factor was then used with the DAVE software to extract the general phonon density of states for the $E_i=29.5$, $59.4$, $124$, and $249$~meV measurements for the integrated ranges of wave-vector transfer $[0.75,3.1]$,$[2,4.5]$, $[2.5,6.5]$, and $[3,8.5]$ {\AA}$^{-1}$ respectively  \cite{dave}.  The individual generalized phonon density of states is shown as a function of energy transfer for four different incident energies in Fig.~\ref{fig:ndos}.  The variation in neutron flux as a function of incident energy was not applied as a normalization for these data.  This accounts for the large variation in the magnitude of the density of states across the measurement configurations, which span an order of magnitude in neutron incident energy.  Each measurement has characteristic features, which we enumerate as we discuss the analysis of these data.

DFT calculations of Cu, Cu$_{2}$S, and LK-99 were performed. The scattering intensity based upon the lattice excitation spectra calculated for the LK-99, Cu$_2$S, and Cu was convoluted individually with the instrumental energy resolution function for the $E_i=29.5$~meV configuration.  The three spectra were added according to the percentages of each compound refined in analyzing the powder diffraction data.  We fit the wave-vector dependent scattering intensity at low energy transfer for the same cut through the data to this calculated resolution convolved scattering intensity, including an overall constant background and an overall multiplicative scaling factor.  The result is shown in Fig.~\ref{fig:ei30cuts}(a) as a solid line that reproduces the quadratic wave-vector dependent measured scattering intensity.  We also fit separately the scattering intensity as a function of energy transfer using the same method for the data in Figs.~\ref{fig:ei30cuts}(b) and (c).  For this case, we include a Gaussian peak at zero energy transfer to account for the elastic scattering. The wave-vector and energy transfer-dependent scattering intensity of the feature at approximately 5 meV energy transfer is well explained from the calculated lattice excitations of the LK-99 and its impurity phases.

We perform a similar fitting of the phonon density of states measurements shown in Fig.~\ref{fig:ndos} to the calculated phonon density of states convolved with the instrumental energy resolution.  We fit simultaneously the four phonon spectra measurements shown in Fig.~\ref{fig:ndos} to the sum of the calculated phonon density of states of LK-99, Cu$_2$S, and Cu with a fixed weighted percentage based upon the refined percentage of these compounds found in the analysis of the diffraction data.  An overall constant background and a multiplicative scaling factor are also included for each of the incident energies in Fig.~\ref{fig:ndos}.  The resulting fit to the phonon density of states is shown as a heavy solid line in the figure.  The individual phonon density of states without the additive background is also shown in this figure separately for the three compounds.  This fit provides an excellent representation of the measured density of states over more than two orders of magnitude of energy transfer.  The bands of calculated phonons agree with calculations of phonon frequencies for LK-99.  \cite{cabezasescares2023theoretical,PhysRevB.108.235127}  There is a larger relative background contribution in the phonon density of states at higher energy transfers.  This is likely due to multiple scattering or multi-phonon effects contributing to these higher energy transfers.  Above $\hbar\omega\approx40$~meV, the phonon density of states is dominated by the contribution from LK-99.  The LK-99 modes in the vicinity of $70$~meV are due to the O-P-O bending modes with motions deforming the PO$_{4}$ tetrahedra. The highest energy mode at approximately 110 meV is due to the stretching modes of the P-O bonds.  One location with a difference in the calculated and measured phonon density of states is in the vicinity of $\hbar\omega=44.6$~meV, Fig.~\ref{fig:ndos}(c).  This mode corresponds to a twisting and scissoring type of motion resulting in squeezing of the PO$_4$ tetrahedra.  The actual energy of this mode appears to be 4 meV higher at $\hbar\omega=48.8$~meV.  We speculate that this difference may be due to the relatively small unit cell used in the calculations.

Earlier, we considered where the superconducting resonance would occur in the spectrum if LK-99 were an unconventional superconductor.  If LK-99 were a normal superconductor, electron-phonon coupling associated with the superconductivity would produce features in the excitation spectrum in the vicinity of the superconducting energy gap, $2\Delta$.  \cite{PhysRevLett.30.214,PhysRevB.12.4899,PRL2Delta_A}   Considering a proposed transition temperature of $T_{C}\approx 400$~K and a range of plus or minus fifty percent deviation from weak, $2\Delta=3.35 k_B T_C$, or strong, $2\Delta=4 k_B T_C$ coupling limits results in a range between approximately 70 and 210 meV over which one would potentially see changes in the phonon spectrum  \cite{superconductivity}.  We observe only modes that are consistent with the calculated phonon density of states in this energy range.  All of the spectral features are well accounted for by the lattice excitations calculated for LK-99 and the Cu and Cu$_2$S impurity phases.\\


\section{Conclusions}

 We present a comprehensive analysis of the crystal structure, encompassing site occupancy, oxygen positions, and polyhedral environments, along with lattice excitations for the copper-substituted lead oxyapatite, Pb$_{10-x}$Cu$_{x}$(PO$_{3.92(4)}$)$_{6}$O$_{0.96(3)}$ (x=0.94(6)). Our structural refinements reveal a favored substitution site of Copper on Pb1 ($6h$) site, which may provide insight into the absence of the Cu-doping-induced structural distortion and superconductivity in LK-99. Our DFT calculations for the Pb$_{10-x}$Cu$_x$(PO$_4$)$_6$O, alongside impurity phases Cu and Cu$_2$S,  predict a phonon spectrum entirely consistent with the measured spectrum.  We find no experimental evidence of static long-range magnetic order, dynamic magnetic fluctuations, superconducting resonances, or superconductivity-correlated electron behavior in the LK-99.


\begin{acknowledgments}
We acknowledge useful discussions with G. Granroth regarding multiphonon backgrounds.  This research used resources at the Spallation Neutron Source, DOE Office of Science User Facilities operated by the Oak Ridge National Laboratory. This research used computing resources made available through the VirtuES project, funded by the Laboratory Directed Research and Development program and Compute and Data Environment for Science (CADES) at ORNL, as well as resources of the National Energy Research Scientific Computing Center (NERSC), a U.S. Department of Energy Office of Science User Facility located at Lawrence Berkeley National Laboratory, operated under Contract No. DE-AC02-05CH11231 using NERSC award ERCAP0024340. This research used the resources of the Center for Nanophase Materials Sciences (CNMS) and Neutron Scattering Division (NSD), which are DOE Office of Science User Facilities. Y.D.G. and Z.Q. M.  acknowledge the support by the US Department of Energy
under grants DE-SC0019068. L.J.M. acknowledges the partial support from NSF
through the Materials Research Science and Engineering Center DMR
2011839 (2020–2026).  
\end{acknowledgments}

\bibliography{LK99bib}

\end{document}